# Comparison of Traditional and Hybrid Time Series Models for Forecasting COVID-19 Cases

Samyak Prajapati[1], Aman Swaraj[2], Ronak Lalwani[1] Akhil Narwal[1], Karan Verma[1]

181210046@nitdelhi.ac.in[1], aman_s@cs.iitr.ac.in[2], 171210048@nitdelhi.ac.in[1], 171210005@nitdelhi.ac.in[1], karanverma@nitdelhi.ac.in[1]

National Institute of Technology, Delhi[1], India
Indian Institute of Technology, Roorkee[2], India

## ABSTRACT

Background: Time series forecasting methods play critical role in estimating the spread of an epidemic. The coronavirus outbreak of December 2019 has already infected millions all over the world and continues to spread on. Many countries have sustained multiple waves of COVID-19 cases and some countries have again started to witness a rise in cases, which is now being referred as the 3rd wave of the pandemic. A thorough analysis of time-series forecasting models is therefore required to equip state authorities and health officials with immediate strategies for future times.

Objective: The aims of the study are three-fold: (a) To model the overall trend of the spread; (b) To generate a short-term forecast of 10 days in countries with extremely high population density like India; (c) To quantitatively determine the algorithm that is best suited for precise modelling of the linear and non-linear features of the time series.

Comparison: Various time-series forecasting models, such as Prophet, Holt-Winters, LSTM, ARIMA, and ARIMA-NARNN have been compared in this study over incidences of Daily Cases and Cumulative Cases of COVID-19.

Result: Single ARIMA performed better than other time series models, however the Hybrid combination of ARIMA and NARNN (ARIMA-NARNN) performed better, where the RMSE was almost 35.3% better than one of the most prevalent method of time-series prediction (Single ARIMA).

Conclusion: The results demonstrated the efficacy of the hybrid implementation of the ARIMA-NARNN model over other forecasting methods such as Prophet, Holt Winters, LSTM, and single ARIMA in encapsulating the linear as well as non-linear patterns of the epidemical datasets.

# 1 INTRODUCTION

The novel coronavirus which first appeared in Wuhan, China in late 2019 has already infected over 257 million people and caused over 5.1 million deaths worldwide [1]. The ground-zero for the zoonotic spillover has been triangulated to the live-food markets of Wuhan, where the virus spread proximally due to direct exposure to animal shedding, bodily fluids, blood, and secretions [2]. In the absence of any tangible treatment, the pandemic has ruptured the concept of normal life while spreading with a rate of 1.8 (in India) [3].

To flatten the pandemic curve, several intervention policies have been implemented in countries all over the world. However, these policies which include mobility and transportation restrictions, have provided temporary relief, but not flattened the curve, which has seen multiple waves of COVID-19 cases as exhibited in Figure 1. The situation has even more degraded in densely populated countries like India and Brazil which can't afford the luxury of lockdown due to socio-economic reasons. Therefore, rapid and predictable up-scaling of the healthcare framework is now most critical towards ensuring the availability of appropriate facilities during these demanding times. In our earlier work we also presented a model for classifying covid-infected X-rays [4].

Each nation now aims at vaccinating their citizens against the virus, but there have been multiple studies claiming that the vaccine elicited immunity is a short-term immunity, and the majority of the populace would require booster shots in the near future. Thus, there lies a sense of uncertainty about the ongoing pandemic and the spread of its contagion.

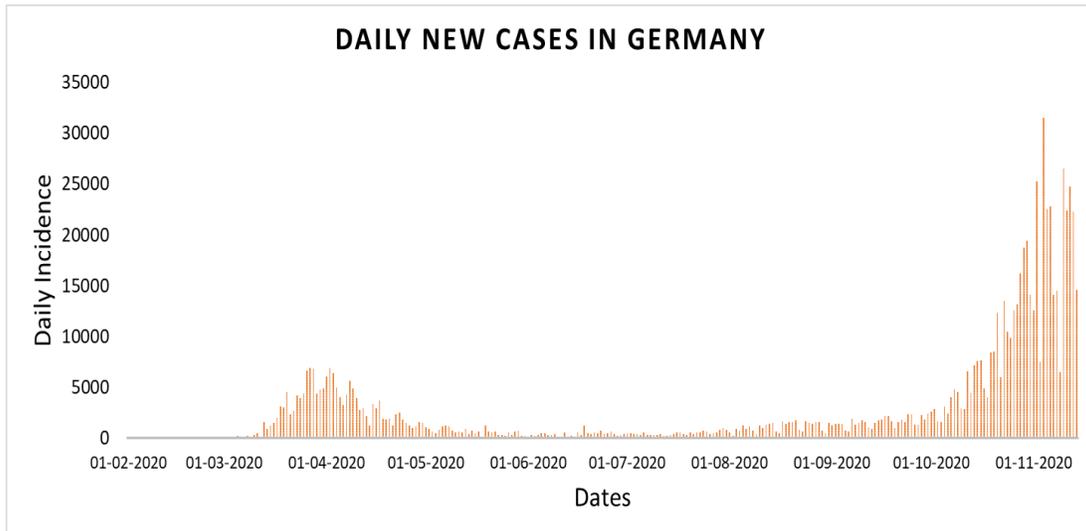

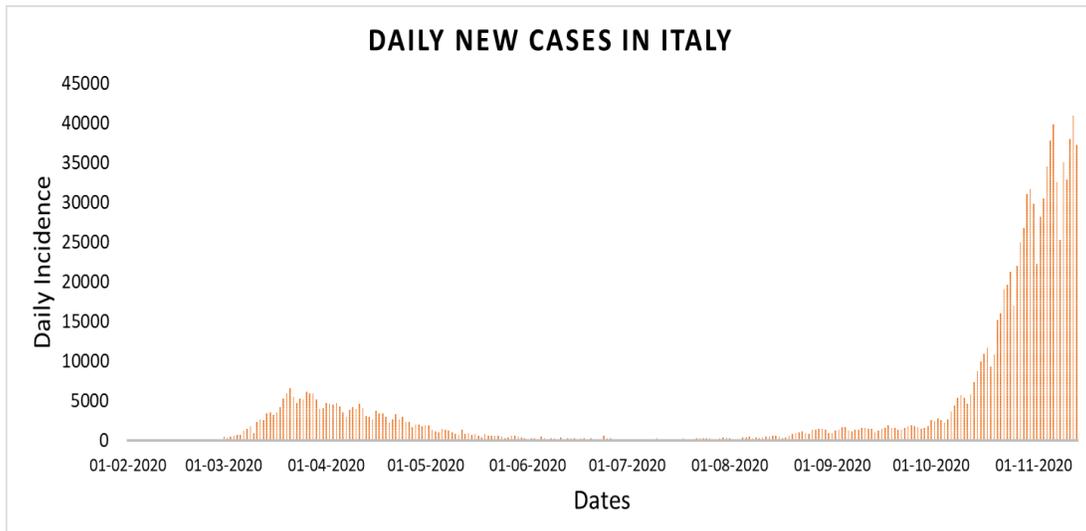

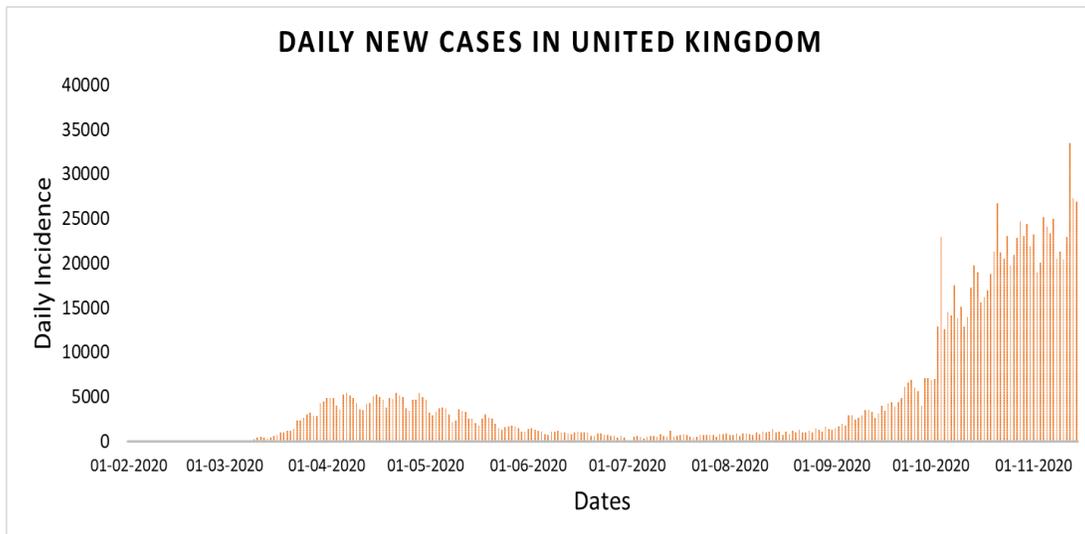

**Figure 1: Second wave of COVID-19 rising in European countries respectively from top: Germany, Italy and UK (February 1 to November 14, 2020)**

Forecasting of epidemics and pandemics has played a key role in curbing the spread of previous epidemics such as Ebola, Influenza etc. [5-11]. Providing insights into the severity of the infection and trend of the outbreak via simplified dashboards, not only helps the general masses to acknowledge the severity of the pandemic but also prompts the state officials to take apt decisions in due time.

One widely used model in discerning the trends of an epidemic is the 'Susceptible–Exposed–Infectious–Resistant' (SEIR) model and researchers have actively employed the model for COVID-19 trend analysis as well [12-20]. While SEIR models are a proven tool in the analysis of these outbreaks, the algorithms and forecasting tools in the domain of Machine Learning and Artificial Intelligence have been considered equally important by researchers for forecasting [21-38].

Some of the commonly used techniques in forecasting data are LSTMs [23-25], Exponential Smoothing [20, 30], Prophet [37, 38], ARIMA [39-41] etc. LSTMs are a form of Recurrent Neural Networks (RNN) with the ability to hold the previous data points for a short period of time which enables the concept of memory in forecasting the spread. Exponential Smoothing works by making use of the lagged values in a weighted fashion; its ease of use and forecasting accuracy are some of the reasons justifying its popularity. Prophet is a fairly new technique developed by Facebook and built on Stan, which is a statistical computation and modelling library; this enables it to be extremely fast in parameter optimization and can easily handle irregular gaps and outliers in the data.

Upon analyzing the related previous works, it is evident that ARIMA usually performs well in forecasting trends where linear patterns dominate the series attributing to its statistical properties and the well-known Box-Jenkins methodology [42]. However, ARIMA model assumes linear correlation structure among the time series values and therefore, it fails to capture the nonlinear patterns accurately. And thus, we propose a hybrid methodology of ARIMA and NARNN (Non-linear Autoregressive Neural Network), where we make use of NARNN for the non-linear patterns and use ARIMA for its forte, of forecasting of linear patterns in a time series.

With this motivation, we analyze and compare multiple time-series forecasting methods namely, Prophet, Holt-Winters, LSTM, ARIMA, and ARIMA-NARNN. Further, we make comparison between all the models and rank them according to their RMSE, MAE and MAPE values.

The rest of the paper is organized in the following sections: In Section 2, we depict the overall flow of the work along with an elaborate explanation of all the chosen models. In Section 3, we present the experiment analysis and results. Section 4 holds a discussion and Section 5 depicts the conclusion.

## 2 METHODOLOGY

This section elaborates on the data collection segment of our work, followed by a short description of the forecasting models that were used. The metrics used to assess the performance of the models are given at the end of this section. The time-series data was fed to all the stated models and their results were compared based on the performance.

## 2.1 Dataset Description

The COVID-19 Data Repository by John Hopkins University's Centre for System Science and Engineering contains the time-series dataset for cumulative count of confirmed cases, reported deaths and recovered cases worldwide [1]. For our study, we chose three countries that were severely affected by COVID-19, respectively the United States, India and Brazil. Since the dataset contained cumulative counts, the data was differenced with its preceding data point to generate a daily incidence time series as well. All the models were trained on three different intervals. The test-train split for each interval is shown in table 1. Last ten days of each respective interval were used for testing purpose.

Table 1. Tabulation of the test-train split of each interval

| Intervals | Training Interval | Testing Interval |
| --- | --- | --- |
| 22JAN - 15MAY | 22nd January – 5th May | 6th May - 15th May |
| 22JAN - 30JUL | 22nd January – 20th July | 21st July - 30th July |
| 22JAN - 10AUG | 22nd January – 31th July | 1st August - 10th August |

The reasoning for 10 days being selected as the test set is because the target of this work was to analyze the performance of forecasting models in terms of their short-term performance during the growing peak of the 'First Wave' of the COVID-19 cases, thus short-term forecasting would be helpful in predicting the spread of COVID-19 and in focusing the attention of the state authorities on a particular region of the country. The models were then ranked accordingly based on Root Mean Square Error (RMSE), Mean Absolute Error (MAE) and Mean Absolute Percentage Error (MAPE). In order to support the results on India, the models were analyzed on USA and Brazil as well (Top 3 countries with the highest incidence of COVID-19).

## 2.2 Models Analyzed

### 2.2.1 ARIMA

Auto-Regressive Integrated Moving Average (ARIMA) was proposed by Box and Jenkins in the 1970s [42] as a model which took varying trends, seasonal changes and random disturbances in account to predict the future values of the series. Due to these reasons, today, it is one of the most popular models that is used for forecasting time-series. It is denoted as ARIMA (p, d, q) where **p** and **q** are the orders of the AR and MA terms of the models respectively, and **d** represents the level of differencing used in the model to achieve stationarity. In a much simpler way, it can be stated that **p** is the number of lagged elements that have an influence on the current element, **d** is the number of times the series is differenced to achieve constant mean and variance and **q** is the number of error terms. It can mathematically be represented as,

$$Y_t = \theta_0 + \phi_1 Y_{t-1} + \phi_2 Y_{t-2} + \cdots + \phi_a Y_{t-a} + E_t - \theta_1 E_{t-1} - \theta_2 E_{t-2} - \cdots - \theta_C E_{t-C} \quad (1)$$

Where $Y_t$ denotes the computed value of forecast at the given time t, $\phi_i$ and $\theta_j$ are the coefficients of the AR and MA models respectively and $E_t$ is the random error occurring at time $t$.

*2.2.2 LSTM*
LSTMs are a form of a recurrent neural network (RNN) and as suggested by their name Long Short Term Memory, they allow for the model to retain information about the data that was previously computed. While most forms of RNNs can utilize the previous data in some form, LSTMs have the intrinsic ability to "store" the data for a short duration. This is achieved by the use of multiple "gates" and by modifying the cell state (Figure 2). Each gate is essentially a function which computes an output determining the way cell state has to be modified.

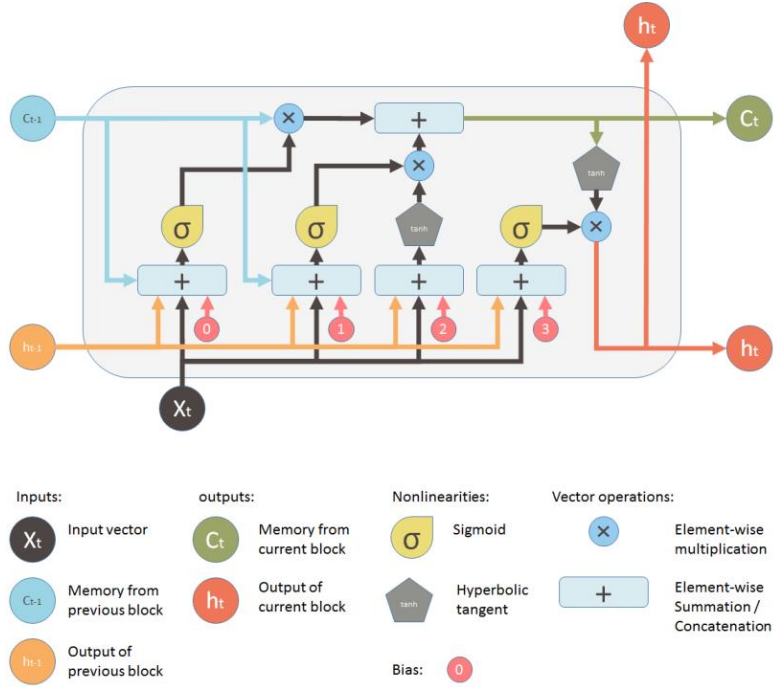

**Figure 2: Pictorial Representation of an LSTM Node [43]**

Each gate can easily be attributed to an activation function, where x is the feature vector, $h_{t-1}$ is the output of cell t-1, $c_{t-1}$ is cell state after cell t-1, $c_t$ is the cell state after cell t and $h_t$ is the output of cell t, thus the resulting computations are,

$$f_t = \sigma(W_f \cdot [h_{t-1}, x_t] + b_f) \quad (2)$$

$$i_t = \sigma(W_i \cdot [h_{t-1}, x_t] + b_i) \quad (3)$$

$$\tilde{c}_t = tanh(W_c \cdot [h_{t-1}, x_t] + b_c) \quad (4)$$

$$c_t = f * c_{t-1} + i_t * \tilde{c}_t \quad (5)$$

$$o_t = \sigma(W_o \cdot [h_{t-1}, x_t] + b_o) \quad (6)$$

$$h_t = o_t * tanh(c_t) \quad (7)$$

*2.2.3 Holt Winters*
Exponential Smoothing is a univariate time-series modelling technique where close attention is paid to the precursing values of the time-series and weights are assigned to them depending on the lag, the combination of the precursing values and their corresponding weights are then factored for prediction of future values. There are mainly three versions of exponential smoothing, which selectively focus on the combinations of Level, Trend and

Seasonality. Single Exponential Smoothing works by modelling the lags of the levels, whereas, Double Exponential Smoothing utilizes levels and trends, and Triple Exponential Smoothing and Holt-Winters Exponential Smoothing incorporates all three elements during its computation. It mainly has 3 parameters,

- $\alpha$: Smoothing factor for the level,
- $\beta$: Smoothing factor for the trend,
- $\gamma$: Smoothing factor for the seasonality.

The mathematical equation for this is described as:

$$F_{i+k} = (L_i + k * B_i) * S_{i+k-m} \qquad (8)$$

Where m is the period length of the seasonal variation, k is the number of steps ahead from any arbitrary step i, and,

$$B_i = \beta * [L_i - L_{i-1}] + (1 - \beta) * B_{i-1} \qquad (9)$$

$$L_i = \alpha * \frac{T_i}{S_{i-m}} + (1 - \alpha) * [L_{i-1} + B_{i-1}] \qquad (10)$$

$$S_i = \gamma * \frac{T_i}{L_i} + (1 - \gamma) * S_{i-m} \qquad (11)$$

*2.2.4 Prophet*

Prophet is an open-source time-series forecasting library developed by Facebook which runs upon Stan, which is a statistical modeling and high-performance statistical computation platform. It is based on a decomposable additive model constituting three major components; trends, seasonality and holidays. The equation for the above can be interpreted as,

$$y(t) = g(t) + s(t) + h(t) + \varepsilon t, \qquad (12)$$

where, $g(t)$ represents the piecewise linear or the logistic growth curve for modelling the non-periodic changes in the time series, $s(t)$ is the periodical changes that occur with seasonality, $h(t)$ includes the effects of holidays (which can be provided by the user) along with schedules that may be irregular in nature and finally, $\varepsilon t$ is the error term which takes in consideration any irregular changes that may not be accommodated by the model.

## 2.3 ARIMA-NARNN Hybrid Model

In our previous work, we have illustrated this point that creation of a hybrid model between ARIMA and a NARNN [44] that can selectively work on a time series by isolating and working on individual areas of strengths (summarized in Figure 3). NARNNs are generally known for their ability to be modelled on non-linear features of a time-series data [45-47]. It works by employing the architecture of recurrent neural networks and uses its embedded memory with feedback connections. This can be exhibited by the mathematical description of the NARNN model,

$$\widehat{Z(t)} = fx(Z(t-1) + Z(t-2) + \cdots + Z(t-n)) \qquad (13)$$

Where $fx$ represents the nonlinear function and the preceding $n$ values of the output determine the future values. In general, a real-world time series contains both, a linear structure and a non-linear component as well. This can be represented as,

$$Z_t = LIN_t + NON_t \qquad (14)$$

Where $Z_t$ is the time-series having a linear component and a non-linear component, which are indicated respectively by $LIN_t$ and $NON_t$.

The first step in this hybridization is to create an ARIMA model with appropriate (p, d, q) parameters. The strength of ARIMA lies in the forecasting of linear dependencies, thus, fitting of the input features into this model will result in the generation of the linear component ($LIN_t$). To isolate the non-linear dependencies, residuals of the ARIMA model must be generated as $RES_t$.

$$RES_t = Z_t - \widehat{LIN_t} \qquad (15)$$

Where $\widehat{LIN}_t$ is the value forecasted by the ARIMA model at a time $t$. By modelling the residuals using ANNs, the non-linear segments can be realized and thus, the residuals are fed into a NARNN model which comprises of n input nodes, modelling it into,

$$R_t = fx\ (R_{t-1}, R_{t-2}, \ldots, R_{t-n}) + \epsilon_t \tag{16}$$

Where, $fx$ constitutes as the non-linear function that is being evaluated by the NARNN model, and the error generated in doing so is represented by $\epsilon_t$. The final equation is then represented by the equation below, where $\hat{Z}_t$ indicates the final forecast of the time-series at time $t$ and $\widehat{NON}_t$ is the residual forecast.

$$Z_t = \widehat{LIN}_t + \widehat{NON}_t \tag{17}$$

## 2.4 Performance Evaluation Measures

Accuracy of a time series model can be evaluated by comparing the predicted values with the actual/true values. There lie several performance measures for this purpose; however, this study employs RMSE, MAE and MAPE. Their mathematical notations are shown below:

$$RMSE = \sqrt{\frac{1}{n}\sum_{t=1}^{n}(Z_t - \hat{Z}_t)^2} \tag{18}$$

$$MAE = \frac{1}{n}\sum_{t=1}^{n}|Z_t - \hat{Z}_t| \tag{19}$$

$$MAPE = \frac{1}{n}\sum_{t=1}^{n}\left|\frac{Z_t - \hat{Z}_t}{Z_t}\right| \tag{20}$$

Here, $n$ stands for the number of data points available, $Z_t$ is the original value at time $t$ and $\hat{Z}_t$ denotes the estimated value at time $t$. Lower values of RMSE, MAE and MAPE indicate the better fitting of the data to the model.

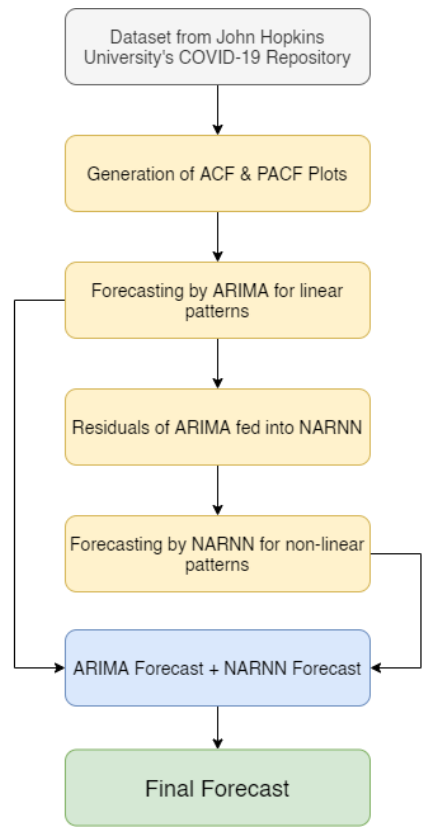

Figure 3: Methodological description of the new hybrid model (Based on [44])

## 3 RESULTS

As described earlier in section 2.1, the models were analyzed on three sets of intervals, (22nd Jan - 15th May and so on, where the last ten days were used for testing), prioritizing India and using the incidence count of the other countries to confirm the observations. In order to attain the best fit of the ARIMA model, the respective parameters p,q and d have to be selected appropriately. And so, first the Augmented Dickey–Fuller (ADF) unit root test was conducted to observe the stationarity of the time-series. Keeping the significance level of 0.05, it was found out that the time-series data was not stationary and needing differencing to achieve stationarity. After differentiating the time-series, ADF tests were repeated to check if stationarity was achieved. After this, ACF and PACF plots of the time series data were generated to identify appropriate AR and MA parameters of the ARIMA model. With p,d,q parameters being identified, the model was fit to the data. AIC and BIC values were used to verify the appropriate fitness of the model to the data; after verification, we achieved the following results as shown in Table 2(a) for India. The actual incidences from each of the models are plotted in Figure 4.

Table 2. (a) Prediction accuracy evaluation for cumulative cases of COVID-19 in India between the 6th and the 15th of May, 21st July and the 30th of July, and Aug 1st and the Aug 10th, 2020.

| Intervals | | Prophet | Holt Winters | LSTM | ARIMA | Hybrid |
|---|---|---|---|---|---|---|
| 6MAY-15MAY | RMSE | 4484.7 | 3556.82 | 7755.93 | 502.30 | 437.30 |
| | MAE | 3206.9 | 279.2 | 5924.5 | 459.6 | 341.8 |
|  | MAPE | 4.330 | 0.436 | 7.871 | 0.718 | 0.523 |
| 21JUL-30JUL | RMSE | 35837.36 | 31493.75 | 132923.72 | 3961.78 | 3119.40 |
|  | MAE | 28300.4 | 24364.1 | 120040.6 | 3150.7 | 2566.7 |
|  | MAPE | 1.946 | 1.662 | 8.474 | 0.246 | 0.197 |
| 1AUG-10AUG | RMSE | 87260.64 | 10152.17 | 49587.15 | 3499.25 | 2825.65 |
|  | MAE | 64841.6 | 8256.5 | 43971.6 | 3041.5 | 2480.5 |
|  | MAPE | 3.118 | 0.420 | 2.179 | 0.156 | 0.127 |

From the above tabulated data in Table 2(a), it is clearly evident that ARIMA performs the best when compared with other popular time-series forecasting methods. Although ARIMA performs good, better still is the hybrid model which is able to map the non-linear components of the forecast as well. To further substantiate the superiority of our Hybrid model over the single ARIMA, we compared the performance of ARIMA and our hybrid model on the same interval on USA and Brazil and the results are tabulated in Table 2(b) and 2(c) and plotted in Figure 5 respectively.

**Table 2. (b) Prediction accuracy evaluation for cumulative cases of COVID -19 in USA between the 6th and the 15th of May, 2020**

| Intervals | | ARIMA | Hybrid |
|---|---|---|---|
| 6 MAY – 15 MAY | RMSE | 5676.96 | 3674.51 |
|  | MAE | 4624.9 | 3009.9 |
|  | MAPE | 0.342 | 0.223 |

**Table 2. (c) Prediction accuracy evaluation for cumulative cases of COVID -19 in Brazil between the 6th and the 15th of May, 2020.**

| Intervals | | ARIMA | Hybrid |
|---|---|---|---|
| 6 MAY – 15 MAY | RMSE | 10062.98 | 7803.56 |
|  | MAE | 8729.90 | 6851.90 |
|  | MAPE | 5.175 | 4.086 |

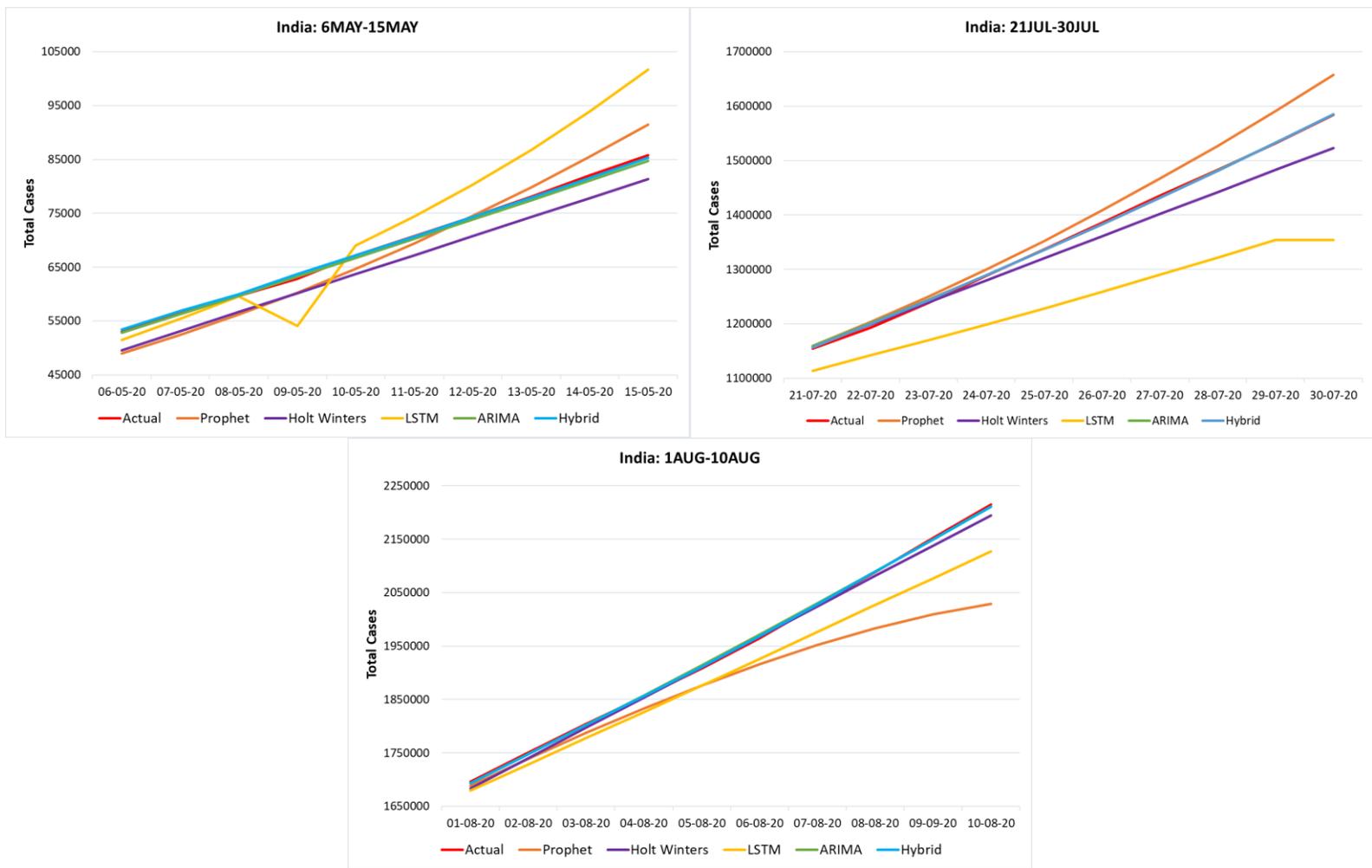

**Figure 4 (a), (b), (c): Graphical illustration of the forecasts for India respectively on 6th May - 15th May, 21st July- 30th July and 1st August- 10th August from the top to bottom**

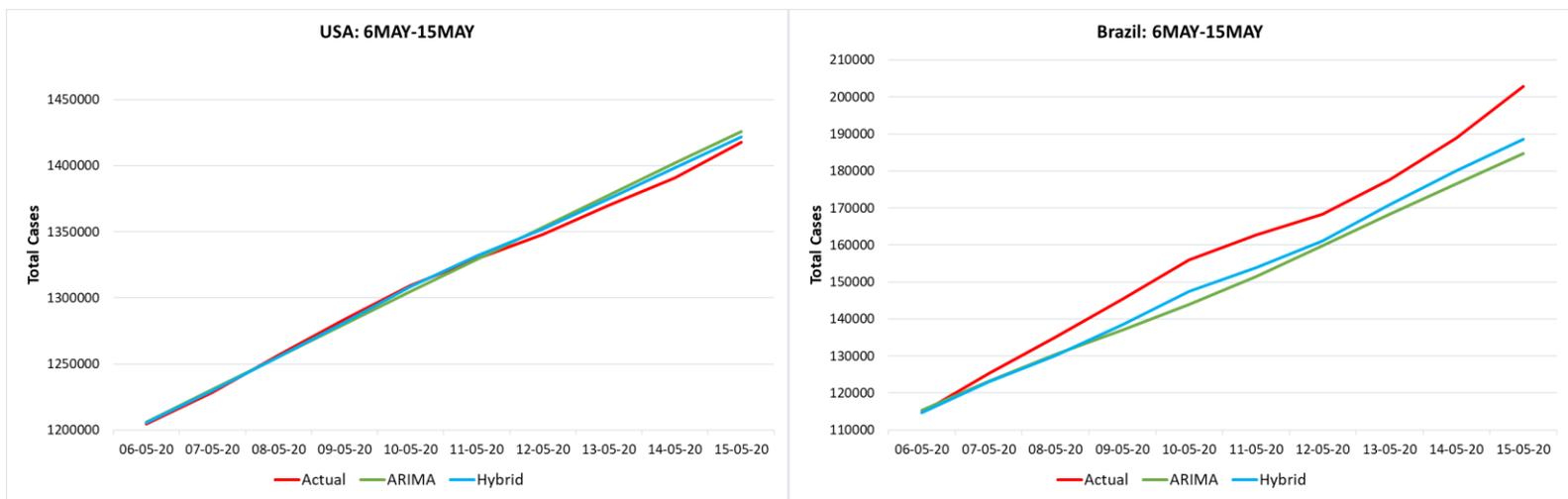

**Figure 5 (a), (b): Graphical Illustration of the forecasts for USA and Brazil respectively from the top to bottom (between 6 - 15th May, 2020)**

It is evident from the aforementioned figures and tables that a hybrid model of ARIMA-NARNN is able to outperform the standard ARIMA by predicting the residuals and working upon them to generate accurate forecast. In order to establish a deeper analysis on India, the hybrid model was further subjected to forecasting daily incidences of COVID-19. Since our prime motive was to compare the efficacy of popular forecasting models in their short-term prediction ability during the first wave of COVID-19 cases, predictions of daily observed cases, daily reported deaths and daily recovered cases were analyzed and tabulated in Table 3 and plotted in Figure 6.

**Table 3. Prediction accuracy evaluation for Daily New Cases of COVID-19, Daily Deaths from COVID-19 and Daily Cases of Recovery from COVID-19 in India between the 6th and the 15th of May, 2020.**

| Type | | ARIMA | Hybrid |
|---|---|---|---|
| Daily New Cases of COVID-19 | RMSE | 329.43 | 275.96 |
| | MAE | 284.90 | 176.92 |
| | MAPE | 7.80 | 4.70 |
| Daily Deaths from COVID-19 | RMSE | 46.39 | 37.79 |
| | MAE | 43.87 | 35.35 |
| | MAPE | 44.02 | 35.32 |
| Daily Cases of Recovery from COVID-19 | RMSE | 198.06 | 177.60 |
| | MAE | 168.14 | 153.34 |
| | MAPE | 10.66 | 9.67 |

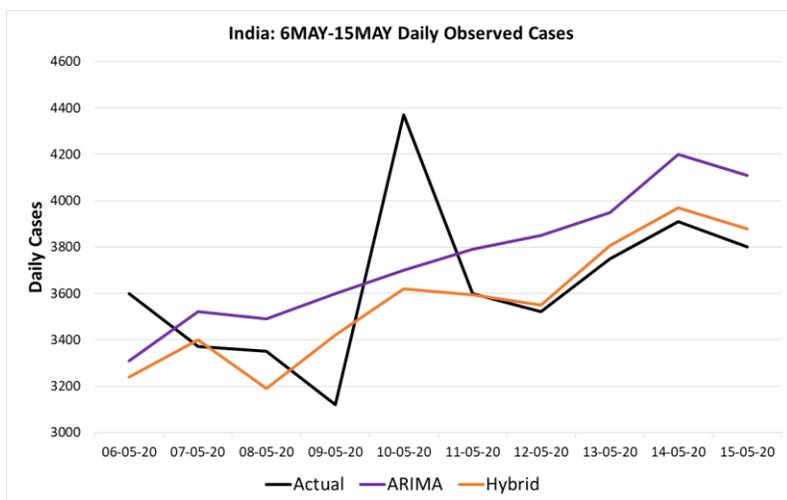
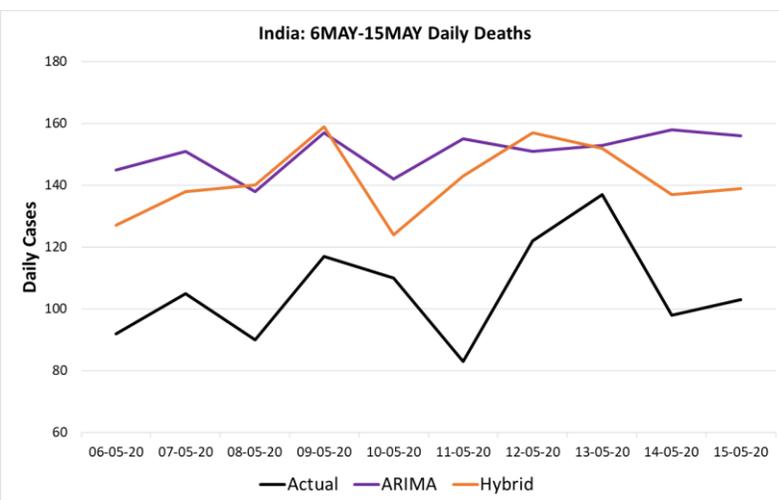

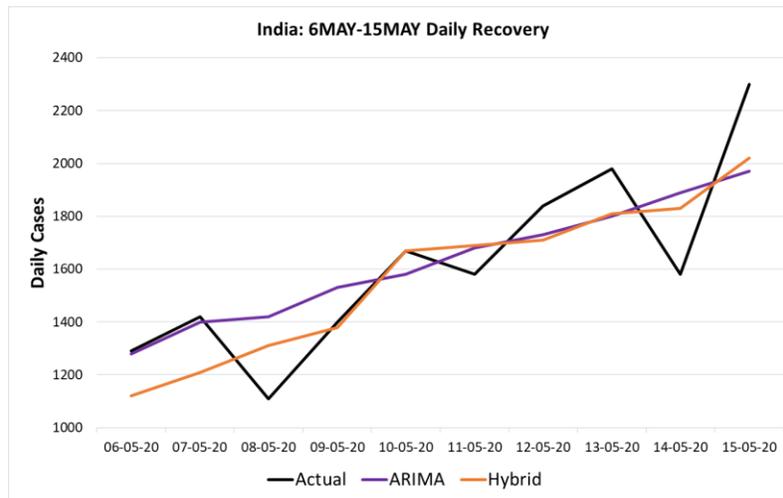

**Figure 6 (a), (b), (c):** Graphical Illustration of the forecasts for India for Daily New Cases of COVID-19, Daily Deaths from COVID-19 and Daily Cases of Recovery from COVID-19 from top to bottom (between 6 - 15th May, 2020)

## 4   DISCUSSION AND CONCLUSION

Our study highlighted the key point of analyzing linear and nonlinear patterns in a time series forecasting model. From Table (2-a, b, c) we see clearly how RMSE value of the hybrid model is minimal when compared to the other stipulated models. This is attributed to the hybrid model having the ability to train itself on the non-linear features of the data.

In the Indian time-series of total cases, there is a notable rise in the non-linear features along with linear features. While ARIMA is able to perform well on just the linear features; the Hybrid model is able analyze the non-linear features as well and is able to substantially improve its performance, thus exhibiting the least RSME amongst the other models.

With most countries hitting their share of the surge of COVID cases, aptly named as the "second-wave", this surge is the result of multiple factors which range majorly from the softening of threat in the mindset of the common folk and the relaxations in the government-imposed policies due to economic slowdowns [48] in several sectors. In such times, where the governing bodies are struggling to stabilize economic recessions, true and precise forecasting of seasonal diseases and the spread of contagions is an ever-growing priority.

Among the chosen time-series forecasting models, ARIMA performed very well, and its performance was then improved by our implementation of the ARIMA-NARNN Hybrid model. While it is important to model the long term spread of contagions, short-term forecasts play a vital role as well in the rapid deployment of resources and manpower especially in developing countries like India and Brazil having dense population and dynamic demographics.

The ever-increasing habitat loss of wildlife leads the animals in search of a new home, this search brings them closer to us; and a consequence of this is that it also exposes us to them. Keeping this in mind, it is plausible that we may get exposed to a lot more zoonotic pathogens in the coming future, and then the only way to circumvent another pandemic is to prepare ourselves in monitoring and curb the spread of infectious diseases.

Practical insights of how the spread of diseases may transpire would lead to the development of an understanding between the policymakers, and hence preferable allocation and management of crucial resources under tight time constraints.

## Conflict of Interest / Competing Interests

The authors declare that they have no conflict of interest. Further, the authors have no relevant financial or non-financial interests to disclose. No funds, grants, or other support was received.

## ACKNOWLEDGMENTS
We thank the Center for Systems Science and Engineering (CSSE) at Johns Hopkins University for providing COVID-19 Data Repository online for open access.

## ABBREVIATIONS
ARIMA: Auto-Regressive Integrated Moving Average
COVID: Corona Virus Disease
LSTM: Long Short Term Memory
MAE: Mean Absolute Error
MAPE: Mean Absolute Percentage Error
NARNN: Non-linear Autoregressive Neural Network
RMSE: Root Mean Square Error
RNN: Recurrent Neural Networks
SEIR: Susceptible–Exposed–Infectious–Resistant